# The role of β pockets resulting from Fe impurities in hydride formation in titanium

Qing Tan [a,b,*], Zhiran Yan [c], Huijun Wang [c], David Dye [b], Stoichko Antonov [a], Baptiste Gault [a,b,*]

[a] Department of Microstructure Physics and Alloy Design, Max-Planck-Institut für Eisenforschung GmbH, Max-Planck-Straße 1, 40237, Düsseldorf, Germany

[b] Department of Materials, Royal School of Mines, Imperial College, Prince Consort Road, London SW7 2BP, United Kingdom

[c] State Key Laboratory for Advanced Metals and Materials, University of Science and Technology Beijing, Beijing 100083, China

[*] Corresponding authors: q.tan@mpie.de, b.gault@mpie.de

**Abstract**

The corrosion potential of commercially pure titanium in NaCl solutions is dramatically affected by trace Fe additions, which cause the appearance of submicron pockets of β phase at grain boundary triple points. Furthermore, the low solubility of hydrogen in hexagonal close-packed α-Ti makes titanium alloys prone to subsequent hydride-associated failures due to stress corrosion cracking. We analyzed α-α and α-β sections of the abutting grain boundary of a β pocket in a Grade 2 CP-Ti, and the α-β phase boundary. Fe and H partition to β and segregate at the grain boundary, but no segregation is seen at the α-β phase boundary. In contrast, a significant Ni (>1 at%) accumulation is observed at the α-β phase boundary. We propose that the β-pockets act as hydrogen traps and facilitate the nucleation and growth of hydrides along grain boundaries in CP-Ti.

Titanium alloys possess excellent fatigue-allowable specific strengths and corrosion resistance, making them highly desirable structural materials for applications in aerospace and biomedical applications [1,2]. In addition, single phase hexagonal close-packed (hcp) α-Ti alloys are among the most corrosion resistant structural materials available for application in aggressive environments [2,3]. Commercially pure titanium (CP-Ti) is a very common and simple class of α-Ti alloys[2], which is categorized into various grades differentiated by their impurity contents – in particular oxygen but also a range of metallic impurities, e.g. Fe and Ni. CP-Ti are corrosion resistant in oxidizing environments, where the protective oxide film remains intact and creates a passive surface, however, their corrosion resistance is limited in strong, highly-reducing acidic media, such as moderately or highly concentrated solutions of HCl, HBr, $H_2SO_4$, and $H_3PO_4$, and in HF solutions at all concentrations, and particularly as temperature increases [4].

The main mode of failure of CP-Ti in aggressive environments is hydrogen embrittlement [5–7]. This degradation of the mechanical properties of CP-Ti, and of many high-strength Ti-alloys, associated to the ingress of hydrogen is well known. H penetrates via absorption and diffusion from a hydrogen-rich environment. Notably, the pitting potential of CP α-Ti alloys in NaCl solutions is strongly affected even by 0.05 wt% Fe as compared to high purity Ti (e.g. IMI110), but is then relatively unaffected by further increases in Fe content within the ASMT Grade 2 CP-Ti allowable range [8]. Historically, this has been attributed to the effect of bulk Fe changes on the stability of the $TiO_2$ scale. Fe stabilizes the high-temperature body-centered cubic (bcc) β-Ti and has little solubility in α-Ti. The high temperature, bcc β-Ti allotrope has a high solubility for H (~ 40 at.%), whilst the low temperature, hcp α-Ti phase has a very limited solubility for H at room temperature (~0.2at.%)[9,10]. Therefore even small amounts of Fe can produce a low volume fraction of retained β-phase in the microstructure in the form of small β pockets, in the range of 100 nm, typically located at triple points across the microstructure. Up until the last decade, these would have been difficult to notice in the SEM or optical microscopy and hence could be easily overlooked. Fe impurities can arise from recycling, from the vacuum distillation apparatus used in the Kroll process, or from the source ores used, although in modern titanium sponge Fe is usually very well controlled. The presence of a low volume fraction of submicron precipitates is quite often important to corrosion and stress corrosion cracking, e.g. MnS in stainless steels, or intermetallic second phase precipitates in Zr alloys, so it is interesting to explore whether triple point β pockets might play a similar role.

A higher content of β-stabilizing impurities, such as Fe and Ni, could allow for a higher density of H trapping sites, which may accelerate hydrogen embrittlement [3]. Since hydrogen is very mobile even at room temperature, it can easily migrate to and embrittle critical microstructural features [11]. Furthermore, α-alloys are highly susceptible to hydrogen embrittlement, which is usually associated to the formation of brittle hydrides [12–14]. Hydrogen concentrations in the α phase quickly become sufficiently high to cause hydride formation [6,15,16]. That is, in comparison with an alloy like equiaxed Ti-6Al-4V with a continuous network of β phase that can allow H to migrate through the structure, CP Ti with triple point β does not allow hydrogen migration at room temperature, but is relatively immune to corrosion.

Elucidating the H behavior within CP-Ti in relation to β-pockets and at interphase and grain boundaries is hence crucial to better understand hydrogen embrittlement. The elemental distribution at and around β-pockets and grain boundaries in CP-Ti has not been studied in detail and the implications for hydride formation in Ti alloys remains unclear. An additional challenge arises from the undesired hydrogen introduction during the specimen preparation for microscopic observations, which cannot be easily avoided. However, previous work has shown that the combination of cryogenic sample preparation via focused ion beam (FIB) and atom probe tomography (APT) can reveal the distribution of H in titanium alloys with sub-nanometer resolution [17–20]. In addition, due to the low solubility of H in α, traditional characterization methods are insufficient to evaluate the H distribution and its effect, and many studies rely on artificially introduced H in the materials via H-charging [21–23]. Although insightful, such conditions can deviate substantially from what the materials face during service.

Here, the β-pockets in a CP-Ti (Grade 2) alloy following mechanical polishing were studied by APT using cryogenic specimen preparation. An α/α grain boundary, α/β grain boundary and α/β phase boundary were characterized and the element distribution at these three types of boundaries elucidated. We reveal H and Fe segregation to the grain boundaries and partitioning to the β phase. These insights suggest that the β-pockets and grain boundaries are the key factors for the formation of hydrides.

Two different Fe-content commercial Grade 2 CP-Ti were sourced from BaoTi. The composition of the alloy was measured by inductively coupled plasma - optical emission spectrometer (ICP-OES), and the H and O contents were measured by an inert gas fusion-thermal conductivity/infrared method. The measured compositions are reported in Table 1.

**Table 1** Chemical composition of the Grade 2 CP-Ti

| Element | Ti | Fe | Si | Cr | Ni | V | C | N | H | O |
|---|---|---|---|---|---|---|---|---|---|---|
| Weight percentage (%) | Bal. | 0.05 | 0.11 | 0.005 | 0.005 | 0.021 | 0.019 | 0.030 | 0.0012 | 0.13 |
| | Bal. | 0.089 | 0.034 | 0.006 | <0.005 | <0.01 | 0.026 | 0.044 | 0.0016 | 0.26 |

Samples were ground to 4000# SiC grid paper and then polished by repeated polishing by colloidal silica in $H_2O_2$ and etching in standard Ti etchant ($HF:HNO_3:H_2O=1:1:10$), to minimize surface hydride formation. The morphology and crystallographic orientations of the microstructure were characterized using back scatted electron (BSE) imaging mode and electron backscattered diffraction (EBSD) analyses on a Zeiss Merlin and JEOL JSM 6400 scanning electron microscope (SEM), respectively.

H was shown to be introduced into titanium alloys during the preparation of specimens for atom probe tomography (APT) [19] by focused ion beam (FIB) milling at room temperature. Here, we used cryo-FIB milling on a gallium-FIB (FEI SEM-FIB Helios nanolab 600i) equipped with a Gatan C1001 cryo-stage using the infrastructure described in [23] to prevent H pick-up [17]. The stage temperature was maintained at -191 °C (82K) over the course of the sharpening process. APT data were acquired on a CAMECA LEAP 5000 XR in high voltage pulsing mode in order to at 50K with a pulsing fraction of 15%, pulsing rate of 200kHz and detection rate of 0.4%. The use of high voltage pulsing leads to better analytical conditions for the analysis of hydrogen in Ti[25]. All data was processed and reconstructed with AP Suite 6.1.

The morphology and grain orientations of a polished CP-Ti surface are shown in Fig 1. Almost all of the hydrides and scratches that were introduced during grinding were removed by the polishing/etching preparation. The grain size of this CP-Ti is 100 μm, and β-pockets about a hundred nanometers in size can be observed located at some and not all triple points, as shown in the inset of Fig 1a. These very small β-phase pockets are formed due to retention of the high temperature β phase during recrystallisation heat treatment, nominally in the single α phase region, due to the presence of β-stabilizing impurities, e.g. Fe, Ni, V, etc. They are also very easily removed by electrochemical polishing, thus using the best possible condition of repeated polishing and etching to get a well-polished surface is extremely important to enable their study.

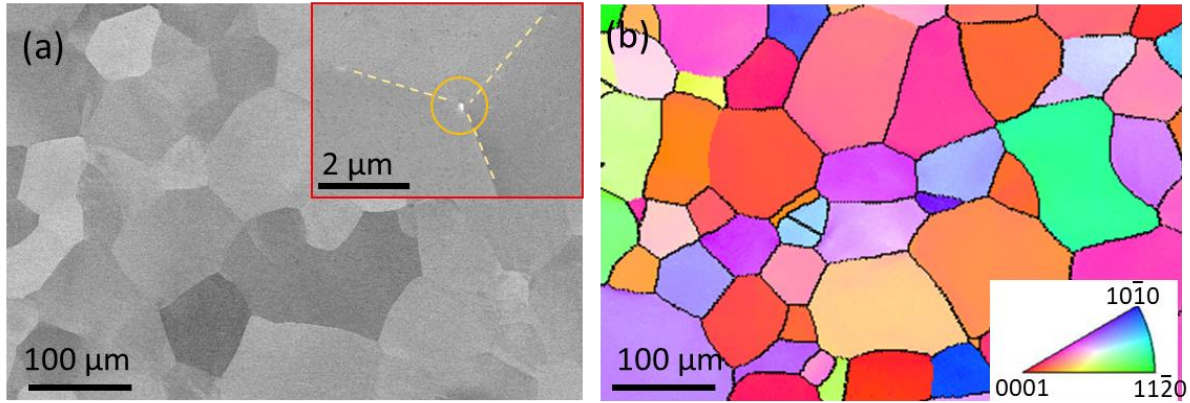

**Figure 1** Back scattered electron image of a well polished CP-Ti. No obvious β-phase can be seen at (a) lower magnification but some very small β-pockets can be observed (bright contrast spot) at triple points at higher magnification (inset). (b) Inverse pole figure map of the received Grade 2 CP-Ti.

APT specimens were prepared from one of these triple point regions of the CP-Ti sample's polished surface, and a reconstruction of a specimen containing a β-pocket is shown in Fig 2a. The region enriched with Fe is the β-pocket where H concentration is also higher, which is expected as Fe and H are both eutectoid β-stabilizers [2]. Importantly, we captured also a section of its adjacent grain and phase boundaries. The boundary denoted in gold in Fig 2a is between two α grains, while the red and olive colored boundary denotations separate the α matrix from the β-pocket. Interestingly, Ni segregates to the boundary denoted by the red line but not the one with olive line, presumably due to a difference between the crystallography of the boundaries. The diffusion coefficients of Fe and Ni in α-Ti are close to each other, and are ~7-8 orders of magnitude greater than those of substitutional impurities, and it has been suggested that these two elements tend to diffuse in an interstitial-like manner in α-Ti [26]. They can hence readily diffuse to the most energetically favorable location during heat treatment and cooling. However, there are no reports on the Ni and Fe partitioning or segregation behavior to interfaces and boundaries. Fig 2c and 2d clearly show that the segregation preference of Ni is grain boundary > β-pocket > α matrix, while that of Fe is β-pocket > grain boundary > α matrix. The different segregation behavior of H, Fe and Ni, appears to be dependent on the different boundary types, as shown in Fig 2a.

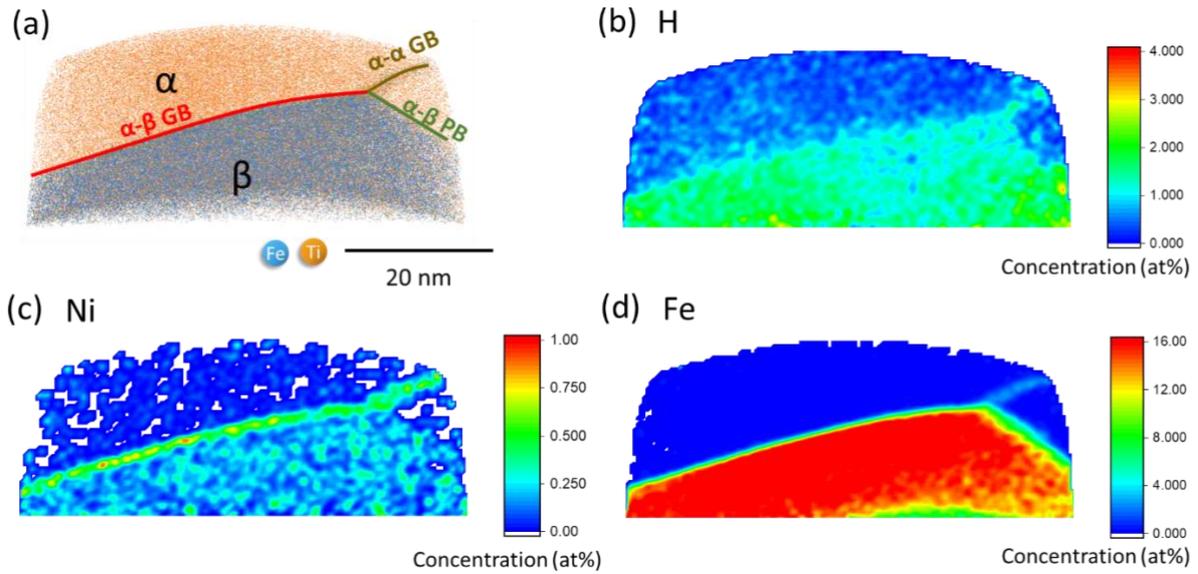

**Figure 2** (a) 3D reconstruction of the Ti (blue) and Fe (brown) ions with boundaries around a β-pocket in CP-Ti, and 2D compositional plots of (b) H (c) Ni and (d) Fe of the same region.

To quantify the elemental distribution to the three different boundaries, cylindrical regions-of-interest were positioned in the dataset (Fig.3a) and 1D composition profiles across the α-α grain boundary (Fig.3b), α-β grain boundary (Fig.3c) and α-β phase boundary (Fig.3d) were calculated. The average H content measured in the α phase was only around 0.4 at.%, which is compatible with what can be expected from the residual gases in the ultra-high vacuum chamber of the atom probe [27]. In the β phase, the H concentration can reach 1.2 - 1.5 at.% and the Fe content is about 18 at.%, consistent with the expected partitioning of H and Fe as β-stabilizing elements. Ni and Cr, which could be picked up in a small amount from the stainless steel reaction vessels during the titanium sponge production process[2], were also detected by APT as well as by ICP-OES and segregate to the β phase. Ni, Fe and H also segregate to the α-α grain boundary, as shown in Fig 3b, which contains 2.5 at.% Fe, 1.0 at.% H and 0.5 at.% Ni. In addition, Ni segregated significantly on the α-β grain boundary (Fig. 3c) with a concentration about 0.8 at.% while no segregation was observed on the α-β phase boundary.

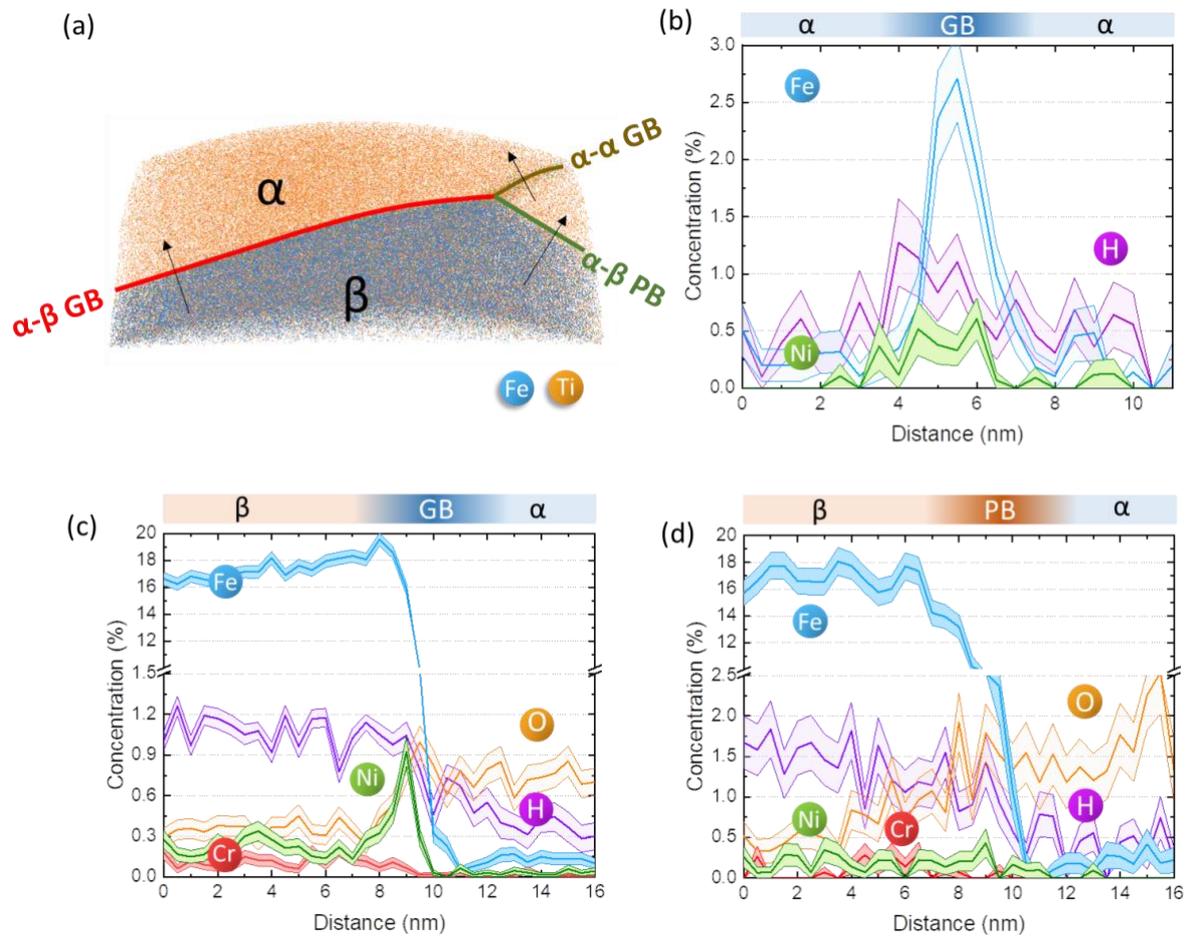

**Figure 3** (a) 3D reconstruction of the Ti (blue) and Fe (brown) ions with three regions of interest perpendicular to each boundary type, and 1D composition profiles across each of the three types of boundaries (a) α-α grain boundary (b) α-β grain boundary (b) α-β phase boundary.

Fig 4a and 4b show BSE micrographs of the two Grade 2 CP Ti alloys with different Fe contents (0.05 wt.% in Fig 4a and 0.089 wt.% in Fig 4b). The Fe-lean sample, Fig 4a, has smaller and fewer β-pockets compared to the Fe-rich alloy, Fig 4b. The brighter dots are β-pockets and the darker needle-like areas are hydrides. Grinding, polishing and etching can all be considered as H-charging processes in Ti alloys[28,29], so we performed manual polishing so as to cause intentional ingress of hydrogen up to a level sufficient to cause the initiation of hydride formation, in order to reveal the initiation sites of the hydrides. In Fig 4a, the large hydrides are almost all co-located with the β-phase and along the grain boundaries. Hydrides can barely be observed inside the grain even around in-grain β phase precipitates (i.e. those formed during cool-down as the solubility for Fe in the α decreases), nor at grain boundaries which are absent of β phase.

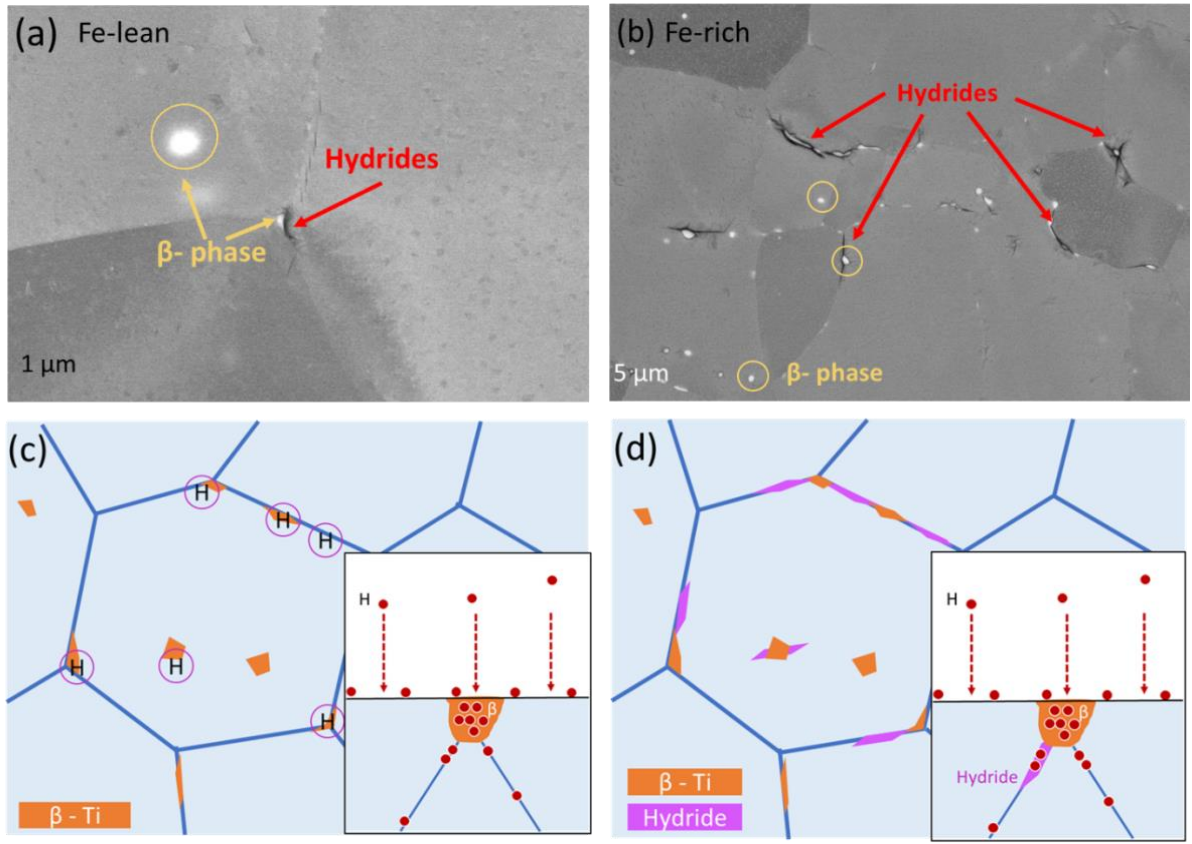

**Figure 4** Not well polished (a) lower Fe content Grade 2 CP Ti with fewer β-pockets and (b) higher Fe content alloy with more β-pockets, both showing the formation of hydrides during polishing. Schematics of (c) hydrogen pick-up and (d) hydride formation in CP-Ti, Insets in c and d are cross-section views.

The α/β phase boundary has been proposed as an effective trapping site for hydrogen in α+β alloys, e.g. in Ti-6Al-4V [30–32]. Meanwhile the α/β boundaries has been shown to be the heterogeneous nucleation sites for hydrides in α+β alloys and are generally considered the main crack propagation pathway for brittle failure[33,34]. In these alloys, the β phase is larger, continuous and with a higher volume fraction compared to the CP-Ti investigated here. The β phase particles in CP-Ti are smaller, discrete and of low volume fraction, but crucially are not connected into a continuous pathway. This enables the study of the β solely as a trap, rather than a combination of fast permeation path and trap. Fig 3 shows no particular H segregation to the α/β phase or grain boundaries, only slight segregation to the α/α grain boundary. Nevertheless, it appears that the hydrides also nucleate from α/α grain boundaries, as shown in Fig 4.

With regards to H generation and ingress, the V rich β-phase in α+β alloys such as Ti-6Al-4V has been shown to be more noble than the α phase[35], with the difference between the two phases increasing with Fe additions[36]. This means that the β phase acts as local cathodic sites, i.e. the location of the hydrogen evolution reaction, thus the ingress of H is through the β phase. Hydrogenation in CP-Ti therefore first proceeds through the β phase, in which the diffusion rate is higher, and remains in solid solution with high solubility, Fig 4c, which is consistent with the reported literature[30,32]. Some hydrogen can also be trapped at the phase and grain boundaries, and the network of grain boundaries can provide faster diffusion paths in-between the β pockets.

As ingress proceeds, the H content can reach the solubility limit of the β phase or, locally, that of the abutting grain boundary. The solubility of H in α is low, the nucleation of a hydride hence ensues as energetically most favorable, Fig 4d. Analogously, in α+β alloys, where a significant amount of continuous β phase is present, hydrogen can be more effectively transported through the β phase, until a high enough concentration is reached to react with the α phase along the α/β boundaries[37]. This is corroborated by a comparison of the hydrogen absorption in duplex and fully lamellar Ti-6Al-4V, where irrespective of the charging conditions, the fully lamellar microstructure always contains higher H content than the duplex microstructure[38], as the α grains can reduce the H absorption sites, viz. the β, and the continuity of the β.

In view of the atom probe data, H and Fe show the same segregation preference and they have a similar interstitial diffusion mechanism [2,26,39] i.e., enriched in the β phase as well as segregated on the α-α grain boundaries, which seem to be the preferred growth pathways of hydrides. Ni has similar segregation preference but is also strongly segregated at the α-β grain boundaries. Whether hydride formation is directly dependent on the presence of Fe still needs to be further clarified, however it seems to be at least indirectly involved by stabilization of the β pockets and increasing their number density, thus increasing the hydride nucleation sites – Fig 4a and b. Phase boundaries between α and β are normally coherent, low energy interfaces, so the driving force for segregation is generally insufficient, even for interstitial elements, such as H. Further studies on unravelling the effective H diffusivity through the different types of boundaries is needed to clarify which boundaries contribute the most to formation of hydrides in CP-Ti.

To summarize, the composition of a triple point β-pocket and its surrounding phase/grain boundaries in Grade 2 CP-Ti was measured by APT from a well-polished surface. Analysis of the element distribution shows that H and Fe exhibit the same partitioning behavior to β, and a segregation to the α/α grain boundary, but not the α/β phase boundary. Ni prefers to segregate to grain boundaries more strongly than to partition to the β phase itself. Based on our observation of the initiation and growth of hydrides from these β-pockets along grain boundaries, we propose that the formation of hydrides in Grade 2 CP-Ti is associated to the ingress of H from discrete β pockets, which are stabilized by impurity Fe, and interaction of H with the α phase at the α/β boundaries. As such, trace β stabilizing elements play a crucial indirect role for hydrogen embrittlement by increasing the fraction of β phase and thus the hydride nucleation sites. Our findings are critical with the increasing importance of recycling of titanium alloys and metallic materials more generally, where there can be concerns around the progressive accumulation of impurities that cannot be removed by vacuum metallurgy, here such as Fe and Ni.

**Acknowledgements**

We acknowledge Uwe Tezins, Christian Broß, and Andreas Sturm for their support to the FIB and APT facilities and Monika Nellessen for her support to the metallography at MPIE. We are grateful to the Max-Planck Society and the BMBF for the funding of the Laplace and the UGSLIT projects respectively, for both instrumentation and personnel. DD, BG, SA and QT are grateful to EPSRC for funding of project EP/T01041X/1. BG and QT are grateful for financial support from the ERC-CoG-SHINE-771602. QT acknowledges the National Natural Science Foundation of China (Grant No. 11805009). SA would like to acknowledge financial support from the Alexander von Humboldt Foundation.